\documentclass[prl,twocolumn,superscriptaddress,floatfix,showpacs]{revtex4}
\usepackage{amsmath,amsfonts,amssymb}
\usepackage{enumerate}
\usepackage{graphicx}
\usepackage{textcomp}
\usepackage{wasysym}
\bibstyle{apsrev.bib}
\begin{document}
\input epsf.sty

\title{Small But Slow World: How Network Topology and Burstiness Slow Down Spreading}

\author{M. Karsai} 
\author{M. Kivel\"a} 
\author{R. K. Pan} 
\author{K. Kaski}
\affiliation{BECS, School of Science and Technology, Aalto University, P.O. Box 12200, FI-00076}
\author{J. Kert\'esz}
\affiliation{BECS, School of Science and Technology, Aalto University, P.O. Box 12200, FI-00076}
\affiliation{Institute of Physics and BME-HAS Cond. Mat. Group, BME, Budapest, Budafoki \'ut 8., H-1111}
\author{A.-L. Barab\'asi}
\affiliation{Institute of Physics and BME-HAS Cond. Mat. Group, BME, Budapest, Budafoki \'ut 8., H-1111}
\affiliation{Center for Complex Networks Research, Northeastern University, Boston, MA 02115}
\author{J. Saram\"aki}
\affiliation{BECS, School of Science and Technology, Aalto University, P.O. Box 12200, FI-00076}
\email{jari.saramaki@tkk.fi}

\date{\today}

\begin{abstract}
While communication networks show the small-world property of short paths, the spreading dynamics in them turns out slow. Here, the time evolution of information propagation is followed through communication networks by using empirical data on contact sequences and the SI model. Introducing null models  where event sequences are appropriately shuffled, we are able to distinguish between the contributions of different impeding effects. The slowing down of spreading is found to be caused mainly by weight-topology correlations and the bursty activity patterns of individuals.
\end{abstract}


\maketitle

Most complex physical, biological and social networks show the {\it small-world} property, where the average shortest path length is strikingly short when compared to the network size  \cite{NWB}. This means that there is at least one short path between any two nodes, which should give rise to rapid transmission of influence. However, dynamic phenomena on networks \cite{Barrat1}, such as spreading of pandemics, electronic viruses, and information, follow their own pathways, which are not necessarily topologically efficient \cite{Holme}. Spreading on real small-world networks turns out to be surprisingly slow, e.g., new infections by a computer virus are reported years after its emergence or the introduction of an anti-virus \cite{SV}. Here we aim at resolving this puzzle. For issues such as strategies and timing of vaccinations, improvement of information diffusion, and the slow decay of prevalence of computer viruses, it is crucial to understand the role of the underlying network and temporal activity patterns in the dynamics of spreading.

The dynamics of spreading is commonly studied with SI, SIR, or SIS models \cite{AM} on static lattices or in mean field, where the dynamics is defined by state changes of individuals between (S)usceptible, (I)nfectious, and (R)ecovered. These models lead to a rapid, exponential growth of prevalence at early stages of spreading, while the dynamics at later stages depend on the model and lattice. For the SI process, the prevalence grows until the whole system reachable from initial conditions is infected, with exponential slowing down towards the end. For the SIR process, competing effects set in and the spreading may remain local or percolate through the system while the SIS process has 
more complex dynamics.
 
While these results capture some of the qualitative features of real-world processes, the heterogeneity of the systems limits their applicability. First, the interactions of real-world systems span networks by broad distributions of node connections and mesoscopic features in the form of communities with dense internal and sparse external connectivity. Second,  interaction intensities vary and are closely coupled to network topology. Third, the daily cycle and bursty character of interaction events give rise to important temporal inhomogeneities. 

Some aspects of these features have already been studied. For static networks, it is known that spatial structure has an effect on epidemics (see, e.g.,~\cite{Keeling,Eames}), and community structure slows down information diffusion due to trapping in dense regions \cite{Lambiotte1,Riitta,Mucha}. There is an intimate relation between inhomogeneous link weights and network topology in social and communication networks \cite{Granovetter,Onnela1}: Links within communities are strong, while links between them are weak. This Granovetter-type structure enhances the trapping effect of the communities, leading to additional slowing down of spreading \cite{Onnela1}. 

The bursty nature of human interactions has received particular interest and it has turned out that the corresponding activity patterns are usually non-Poissonian, often power-law correlated (see \cite{BarabasiBursts}). The effect of bursty dynamics on spreading has been approached using empirical data together with approximate analytical models~\cite{Vasquez1,Iribarren}. In Ref.~\cite{Vasquez1}, computer worm spreading was studied using email logs and the SI model, and it was found that the non-Poissonian inter-event time distribution leads to slow spreading in the late stages of the process. Slow spreading was also observed in Ref.~\cite{Iribarren}, where an Internet viral marketing experiment was carried out and modeled as a branching process in the non-percolating regime. It was also argued that on the contrary, in the percolating regime, broad inter-event time distributions should give rise to faster spreading.

In this Letter, we study the problem of spreading dynamics in its full complexity, using time-stamped event data on human communication networks and the SI model. We apply proper null models on the event sequences and show that spreading is slowed down due to simultaneous effects of structural and temporal correlations.

For the event sequences, we have used the following data: a)  Mobile phone data from a European operator (national market share $\sim20$\%) with $\sim325$ million time-stamped voice call records over a period of 120 days. We have only retained links with bidirectional calls within the largest connected component (LCC) of the aggregated call network (MCN), yielding $N=4.6\times 10^6$ nodes, $L=9\times 10^6$ links, and $306\times 10^6$ calls. We define link weights as the number of calls between two users. The network is sparse (average degree $\left\langle k\right\rangle=3.96$) showing small world property with an average shortest path length of  $\langle l \rangle=12.31$; b) Mobile call data from the Reality Mining project \cite{Eagle} (RM), where the LCC consists of 59 users and 93 edges with 2293 calls over $\sim9$ months; c) email logs \cite{enron} forming a network with the LCC having 2993 nodes and 28843 edges for 202687 events over 83 days. Here communications are directed and thus the nodes belong to the strongly connected component (SCC) where all nodes are reachable from each other, or the IN- or the OUT-component. 
 
We study the SI spreading dynamics with simulations using the event sequences so that an infected individual infects a susceptible one at time $t$, if there is an event between them.
For the events, we use records of the times and participants of calls, and the times and addresses of emails. Calls are one-to-one communication and enable {\em bidirectional} exchange of information, while emails may have multiple addresses and the information flow is {\em directed}. Hence for calls, if either participant is infected he/she infects the susceptible one, whereas for emails, transmission is from the sender to the recipient(s). We initiate simulations by infecting a randomly chosen node at a randomly chosen event with the  spreading quantity (information, rumor, or virus) and set all other nodes susceptible. Then the spreading dynamics is simulated by using temporally periodic boundary conditions ({i.e.}, repeating the event sequence) until the set of reachable nodes is exhausted. We record the prevalence, {i.e.},~the fraction of infected nodes  $\left<I(t)\right>/N$ as a function of time averaging over $10^3$ initial conditions and the time to full prevalence $t_f$. For the email network, we start the spreading process from a node in IN or SCC and iterate the process until all nodes in SCC and OUT are infected.

\begin{table}[b!]
\begin{center}
\begin{tabular}{|l||c|c|c|c|c|c|}
\hline
EVENT SEQUENCE & D & C & W & B & E \\ \hline\hline

Original &  \checkmark & \checkmark  & \checkmark & \checkmark & \checkmark  \\ \hline
Equal-weight link-sequence shuffled  & \checkmark &  \checkmark & \checkmark & \checkmark & \\ \hline
Link-sequence shuffled & \checkmark &  \checkmark & & \checkmark &  \\ \hline
Time shuffled & \checkmark & \checkmark  & \checkmark & &  \\ \hline
Configuration model & \checkmark  & & & & \\ \hline
\end{tabular}
\end{center}
\caption{Correlations retained in different null models. D: daily pattern, C: community structure, W: weight-topology correlations, B: bursty single-edge dynamics, E: event-event correlations between edges.}
\label{table1}
\end{table}

To gain insight into the effects of different correlations, we employ null models where the original event sequences are randomized. These are defined so that in each null model, some of the correlations are separately destroyed: community structure (C), weight-topology correlations (W), bursty event dynamics on single links (B), and event-event correlations between links (E). In addition, the overall event frequencies follow a daily pattern (D), with decreased night-time activity and some day-time peaks (see inset in Fig.~\ref{fig:Poisson})  The null models are as follows, with the letters indicating retained correlations (Table~\ref{table1}):

\noindent
-- DCWB \emph{(equal-weight link-sequence shuffled)}: Whole single-link event sequences are randomly exchanged between links having the same number of events. Temporal correlations between links are destroyed. (For large weights we did binning with 2-3 weight values.)

\noindent
-- DCB \emph{(link-sequence shuffled)}: Whole single-link event sequences are randomly exchanged between randomly chosen links. Event-event and weight-topology correlations are destroyed.

\noindent
-- DCW \emph{(time-shuffled)}: Time stamps of the whole original event sequence are randomly reshuffled. Temporal correlations are destroyed.

\noindent
--  D \emph{(configuration model)}: The original aggregated network is rewired according to the configuration model, where the degree distribution of the nodes and connectedness are maintained but the topology is uncorrelated. Then, original single-link event sequences are randomly placed on the links, and time shuffling as above is performed. All correlations except seasonalities like the daily cycle are destroyed.

\begin{figure}[!bl]
  \begin{center}
     \includegraphics[width=7.4cm,angle=0]{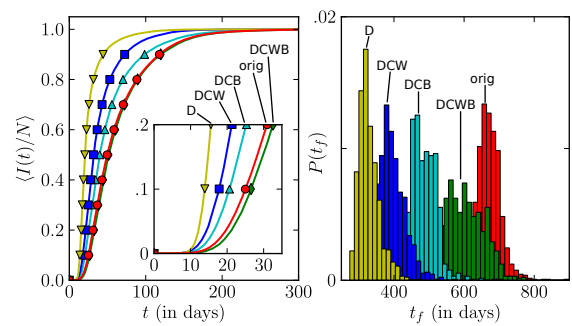}
  \end{center}
  \caption{(color online) (Left) Fraction of infected nodes $\left<I(t)/N\right>$ as a function of time for the original event sequence (\textopenbullet) and null models:  equal-weight link-sequence shuffled DCWB ($\lozenge$), link-sequence shuffled DCB ($\vartriangle$), time-shuffled DCW ($\square$) and configuration model D ($\triangledown$). Inset: $\left<I(t)/N\right>$ for the early stages, illustrating differences in the times to reach $\left<I(t)/N\right>=20\%$. (Right) Distribution of full prevalence times $P(t_f)$ due to randomness in initial conditions.}
   \label{fig1}
\end{figure}

\begin{figure}[!htl]
  \begin{center}
     \includegraphics[width=7.0cm,angle=0]{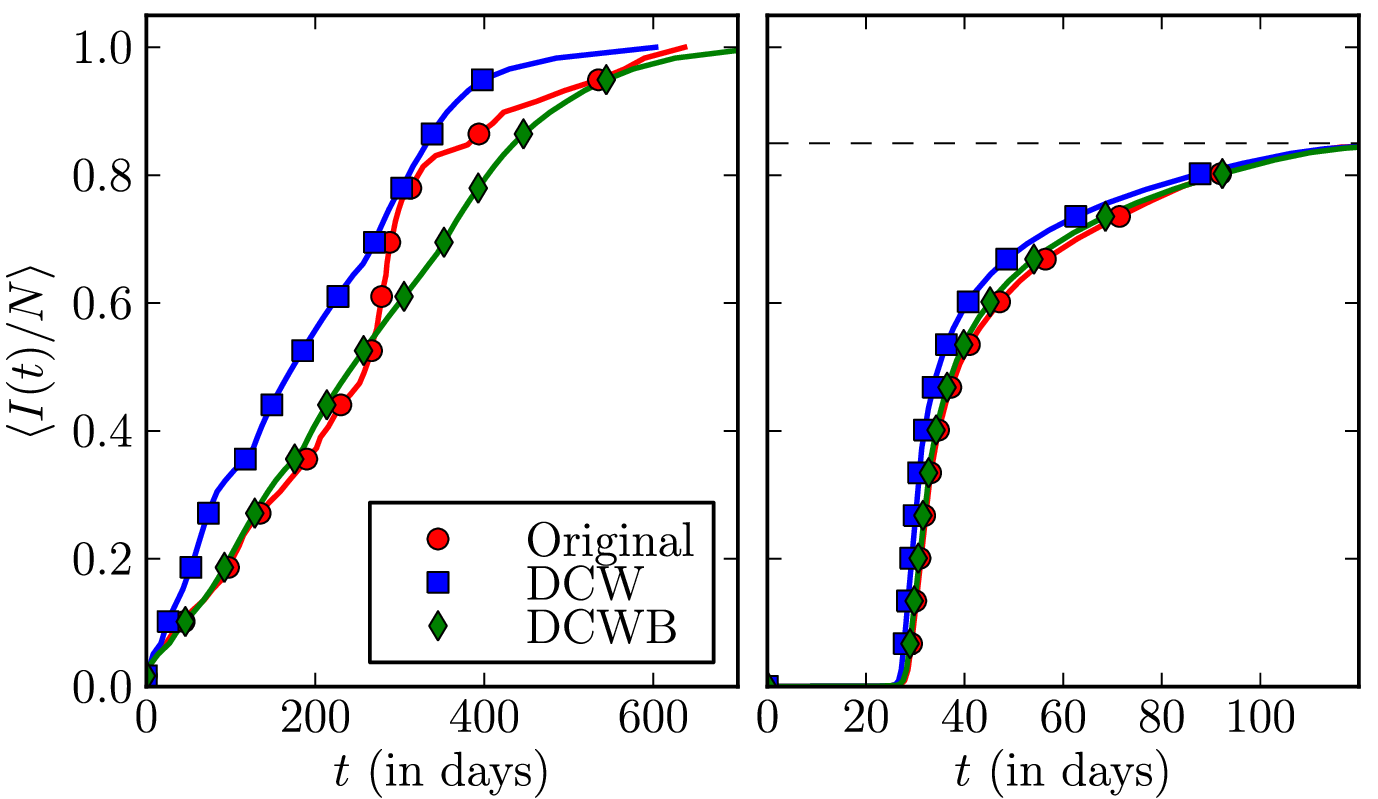}
  \end{center}
  \caption{(color online) Spreading dynamics in the Reality Mining (left) and email networks (right), for the original event sequence (\textopenbullet) and null models: DCW ($\square$) and DCWB ($\lozenge$). In the email network, the spreading process is directed. The maximum prevalence is limited to the total fraction of the SCC and the OUT component ($\sim85$\%).}
   \label{fig3}
\end{figure}

Fig.~\ref{fig1} displays the results for the MCN.  In all cases the spreading is slow, with full prevalence times $t_f$ of the order of several hundred days. It is clear that both topological and temporal correlations slow down the spreading. It is the fastest when all correlations except the daily patterns are destroyed (configuration  model, D). Switching on the community structure and associated weight-topology correlations (DCW) slows down the spreading strongly, as expected because of the bottleneck caused by weak links between communities and the broad distribution of link weights~\cite{Onnela1,Onnela2} . However, comparing this with the DCB null model indicates that bursty single-edge dynamics (B) has an even stronger slowing-down effect than weight-topology correlations (W). Finally, including all except event-event correlations (DCWB) gives rise to spreading dynamics very close to the original event sequence (DCWBE). Here, for early times, DCWB spreading is slightly slower than the original one. The left panel inset shows quantitative differences in the times to 20\% prevalence. It also indicates that temporal correlations (E) between adjacent edges have initially a minor accelerating effect. This can be attributed to the easy reachability of the members within the community where the spreading begins. However, for long times, bottlenecks appear, and event-event correlations slow the process down. Note that the initial conditions have an effect on the duration of the process, reflected in the distributions in the right panel of Fig.~\ref{fig1} (the SI process itself is deterministic). However, the overall shape of the dynamics and the effects of correlations are consistent for individual runs too.

\begin{figure}[!htl]
  \begin{center}
     \includegraphics[width=7.0cm,angle=0]{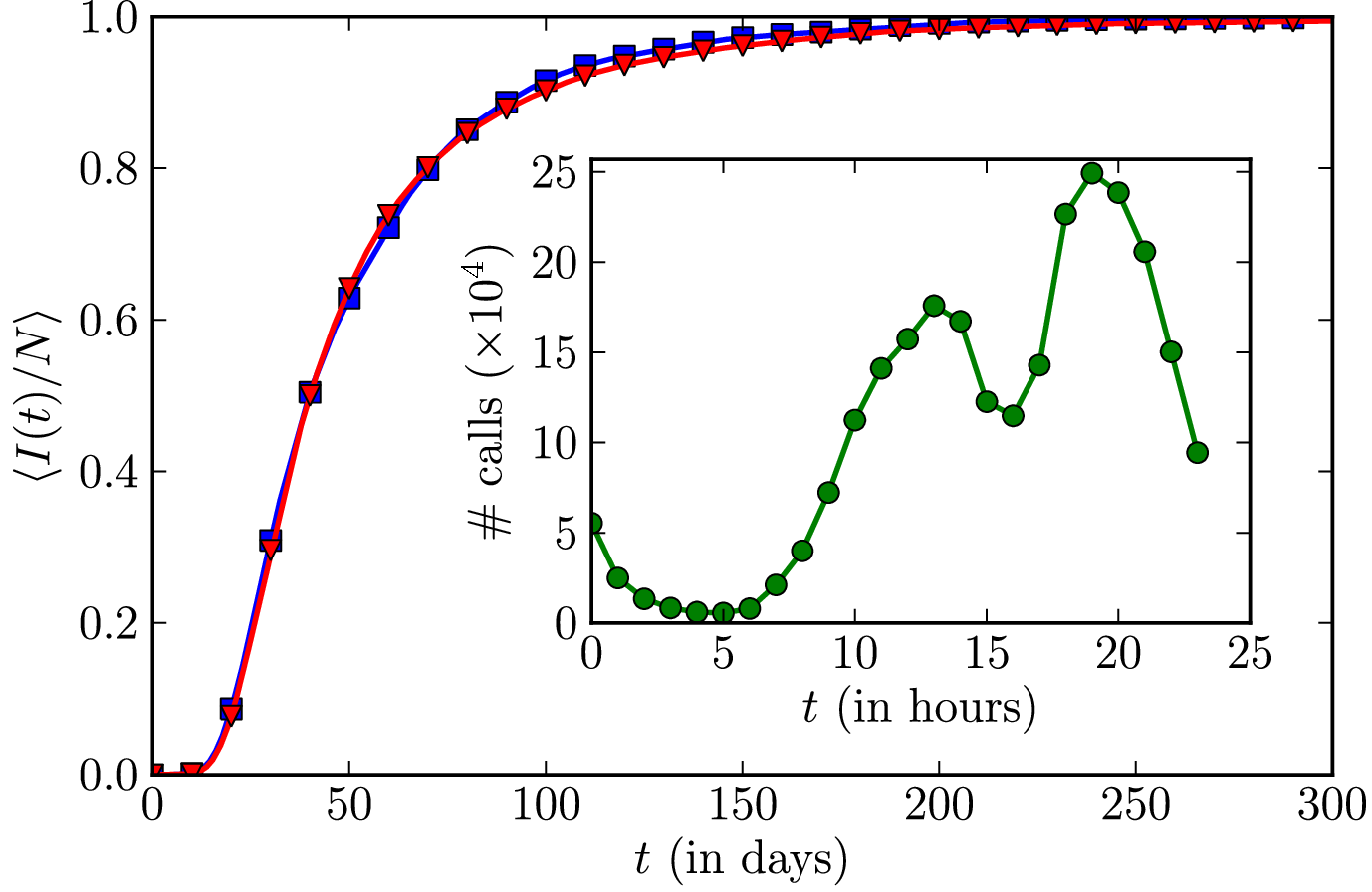}
  \end{center}
  \caption{(color online) Spreading dynamics as obtained from a Poissonian event-generating model on the aggregated MCN, with daily pattern ($\square$) and without ($\triangledown$). Link weights were taken into account and the curve with the daily pattern is comparable with the DCW null model. Inset: the average daily pattern as observed for the MCN event sequence with binning by the hour. The continuous line is to guide the eye.}
   \label{fig:Poisson}
\end{figure}

Results for the Reality Mining mobile call network and for the email logs are shown in Fig.~\ref{fig3}, with the DCW and DCWB null models; the outcome is qualitatively similar with that of MCN. However, there are certain differences. In the small and sparse RM network, successive calls to many people within a short time period by a hub give rise to a steep prevalence rise. Such behavior is a one-off event and the effect is destroyed in the null models. In the email network, very high-degree hubs sending frequent emails give rise to rapid spreading once they are reached. This effect is conserved in the null models.

The daily activity pattern, \emph{i.e.}~variation in overall communication frequency by the hour, is retained in every null model that is based on randomizing the original event sequence. In~\cite {Malmgren}, it was suggested that natural periodicities, such as the daily cycle, are responsible for the fat-tailed waiting time distributions. In order to evaluate the impact of the daily pattern on the spreading speed, we carried out simulations where the aggregated MCN was used as the lattice. Events were generated on its links by two Poisson processes that conserve link weights: a homogeneous Poisson process, and a process whose instantaneous rate follows the daily pattern as calculated from the call statistics on hourly basis (see inset in Fig.~\ref{fig:Poisson}). The SI dynamics for both cases are shown in Fig.~\ref{fig:Poisson}. The difference between the two curves is negligible, demonstrating that the daily pattern has only a minor impact on the spreading speed. This, together with the observation that temporal correlations do have a significant decelerating effect on spreading strongly indicates that there are important, non-Poissonian correlations in the system beside the daily type cycles. 

\begin{figure}[!tl]
  \begin{center}
     \includegraphics[width=7.0cm,angle=0]{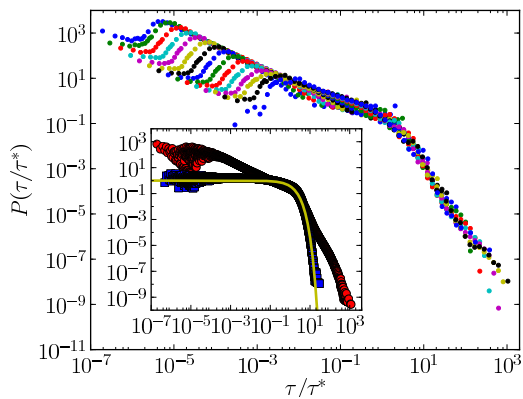}
  \end{center}
  \caption{(color online) Scaled inter-event time distributions for the MCN. Edges were log-binned by weight and for every second bin the inter-event time distribution of the events occurring in the corresponding bin is shown, scaled by the average inter-event time of that bin $\tau^*$. Inset: scaled inter-event time distributions for the original (\textopenbullet) and for the time-shuffled events ($\square$). An exponential density distribution with average value of $1$ is shown as a light (yellow) line.}
   \label{fig:distrib}
\end{figure}

The non-Poissonian, bursty character of event sequences is clearly demonstrated by the fat-tailed distribution of single-link inter-event times for the MCN, as seen in Fig.~\ref{fig:distrib}. In order to exclude the possibility that the fat tail in the inter-event time distribution is only due to the broad weight distribution as suggested in \cite{Malmgren}, we calculated the distributions for binned weights and obtained a satisfactory scaling with the average inter-event time, similarly to \cite{Candia}.  We find that the distribution can be fitted by a power law with an exponent $~0.7$ over 3.5 decades, followed by a fast decay. The scaling breaks down for small inter-event times, where a peak in the distribution at $\sim20$ seconds is found. This peak is due to event correlations between links. The power law indicates the non-Poissonian, bursty character of the events. Both the characteristics vanish for the time-shuffled null model DCW, and the inter-event time is well described by an exponential function (see inset of Fig.~\ref{fig:distrib}), i.e., the process is Poissonian.

The effect of burstiness on the spreading speed can be easily demonstrated with the following single-link calculation. Let us denote the average time for the infection to spread through a link (the residual waiting time) by $\left< \tau_R \right> $, and assume that one of the nodes gets infected at a uniformly chosen random time. Similarly to Iribarren \emph{et al.} \cite{Iribarren} and Vazquez \emph{et al.}, \cite{Vasquez1} we calculate $\left< \tau_R \right> $ for a given inter-event time distribution $P(\tau)$. For simplicity, we consider how the burstiness introduced by a continuous power-law distribution of inter-event times $P(\tau)\sim \tau^{-\alpha}$ affects the average infection times when compared to a Poisson process. If we fix the average inter-event time (and thus the number of events for a long observation period), the ratio of average infection times is $r=\left< \tau_{R,\mathrm{powerlaw}} \right> /\left< \tau_{R,\mathrm{poisson}} \right> = \frac{(\alpha -2)^2}{2(\alpha -1)(\alpha -3)}$ for $\alpha > 3$. Now $r$ is decreasing with $\alpha$, $r<1$ when $\alpha>2+\sqrt{2}\approx 3.4$, and $r$ goes to infinity at $\alpha=3$. This indicates that the burstiness characterized by power law distributions with slow decay has a decelerating effect on spreading with respect to the Poisson process with the same mean. However, if the decay is fast enough, i.e., the second moment of the power law distribution is smaller than that of the Poisson distribution, we see acceleration. This mean field type of reasoning has its limitations. Nevertheless it illustrates the mechanisms of slowing down because of bursts: the residual waiting time increases because the chance for long waiting times after getting infected increases.

In conclusion, we have studied the effects of different topological and temporal correlations on spreading in complex communication networks. Using time-stamped event data and appropriately prepared null models we  have managed to quantitatively distinguish between different contributions to the slowing down of spreading. We have shown that the main contributions are (i) the community structure and its correlation with link weights and (ii) the inhomogeneous and bursty activity patterns on the links. Somewhat surprisingly, the daily pattern and event correlations between links seem to play only a minor role in the overall spreading speed. Finally, we believe that our null models can be generally applied to investigate the effects of temporal and structural correlations on dynamic processes on networks.

{\bf Acknowledgement} Financial support from EU's 7$^{th}$ Framework Program's FET-Open to ICTeCollective project no.~238597 and by the Academy of Finland, the Finnish Center of Excellence program 2006-2011, project no.~129670, as well as  by OTKA K60456 and TEKES (FiDiPro) are gratefully acknowledged.

\end{document}